\documentclass{ws-procs975x65}

\def\citep{\cite}

\def\citet{\cite}

\def\msol{M_\odot}

\begin{document}

\title{EOS OF DENSE MATTER\\ AND\\ FAST ROTATION OF NEUTRON STARS}

\author{J.L. ZDUNIK$^*$, P. HAENSEL, M. BEJGER }

\address{Nicolaus Copernicus Astronomical Center\\ Polish
           Academy of Sciences\\ Bartycka 18, PL-00-716 Warszawa,
           Poland\\
$^*$E-mail: jlz@camk.edu.pl}

\author{E. GOURGOULHON}

\address{LUTh, UMR 8102 du CNRS, Pl. Jules Janssen,\\ 92195
            Meudon, France}

\begin{abstract}
Recent observations of XTE J1739-285 suggest that
it contains a neutron star rotating at 1122 Hz\cite{Kaaret2007}.
Such rotational frequency would be the first for which the
effects of rotation are significant.
We study the consequences of very fast rotating neutron stars
for the potentially observable quantities as stellar mass
and pulsar period.
\end{abstract}


\bodymatter
\section{Introduction}
\label{sect:introd}

Neutron stars with their very strong gravity can be very fast
rotators. Theoretical studies show that they could rotate at
sub-millisecond periods, i.e., at frequency $f=1/{\rm
period}>$1000 Hz\cite{CST1994,Salgado1994}. The first
millisecond pulsar B1937+21, rotating at $f=641$
Hz\cite{Backer1982}, remained the most rapid one during 24
years after its detection. In January 2006, discovery of a
more rapid pulsar J1748-2446ad rotating at $f=716$ Hz
 was announced \citep{Hessels2006}. However, such  sub-kHz
 frequencies are still too low to
significantly affect the  structure of neutron stars with
$M>1M_\odot$ \cite{STW1983}. Actually, they belong to a {\it
slow rotation} regime, because their $f$ is significantly
smaller than the mass shedding (Keplerian) frequency $f_{\rm
K}$. Effects of rotation on neutron star structure are then
$\propto (f/f_{\rm K})^2\ll 1$. Rapid rotation regime for
$M>1M_\odot$ requires submillisecond pulsars with supra-kHz
frequencies $f>1000$ Hz.

Very recently Kaaret et al.\cite{Kaaret2007} reported a
discovery of oscillation frequency $f=1122$ Hz in an X-ray
burst from the X-ray transient,  XTE J1739-285. According to
Kaaret et al.\cite{Kaaret2007} "this oscillation frequency
suggests that XTE J1739-285 contains the fastest rotating
neutron star yet found". If confirmed, this would be the first
detection of a sub-millisecond pulsar (discovery of a 0.5 ms
pulsar in SN1987A remnant announced in January 1989 was
withdrawn one year later).

Rotation at $f>1000$ Hz is sensitive to the stellar mass and
to the equation of state (EOS).  Hydrostatic, stationary
configurations of neutron stars rotating at given rotation
frequency $f$ form a one-parameter family, labeled by the
central density. This family - a curve in the mass -
equatorial radius plane - is limited by two instabilities. On
the high central density side, it is instability with respect
to axi-symmetric perturbations, making the star collapse into
a Kerr black hole. The low central density boundary results
from the mass shedding from the equator.  In the present paper
we show how rotation at $f>1000$ Hz is sensitive to the EOS,
and what constraints on the EOS of neutron stars result from
future  observations of stably rotating sub-millisecond
pulsars.

\section{Method}

\begin{figure}[h]
\centering
\resizebox{3in}{!}{\includegraphics[]{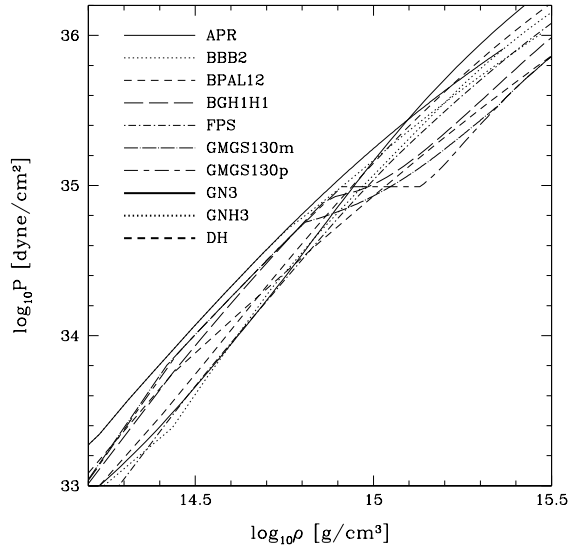}}
\caption{The equations of state used in the paper}
\label{fig:eos}
\end{figure}

We studied the properties of fast rotating neutron stars for a broad set of
the models of dense matter. The set of equations of state (EOSs) considered in the
paper is presented in Fig.~\ref{fig:eos}.

Out of ten EOSs of neutron star matter, two were chosen to represent
a soft (BPAL12) and stiff (GN3) extreme case. These two extreme
EOSs should  not  be considered as ``realistic'', but they are used just to
``bound'' the neutron star models from the soft and the stiff side.

We consider four EOSs based on realistic models
involving only nucleons (FPS, BBB2, DH, APR),
and four  EOSs  softened at high
 density either by the appearance of hyperons (GNH3, BGN1H1),
 or a phase transition (GMGSm, GMGSp).
The softening of the matter in latter case is clearly visible in Fig.~\ref{fig:eos} at
pressure $P\sim 10^{35}~{\rm dyn/cm^2}$.

EOSs GMGSp and GMGSm describe nucleon matter with a first
order phase transition due to kaon condensation. In both cases
the hadronic
Lagrangian is the same. However, to get GMGSp we
assumed that the phase transition takes place between two pure
phases and is accompanied by a density jump.
Assuming on the contrary that the transition occurs via a mixed state of two
phases, we get EOS GMGSm. This last situation prevails when the
surface tension between the two phases is below a certain
critical value.

The  stationary configurations of rigidly rotating neutron
stars have been computed in the framework of general relativity
by solving the Einstein equations for stationary axi-symmetric
spacetime\cite{BGSM1993,GourgSS1999}. The
numerical computations have been performed  using  the {\tt
Lorene/Codes/Rot\_star/rotstar} code from the LORENE library
({\tt http://www.lorene.obspm.fr}). One-parameter
families of stationary 2-D configurations were
calculated for ten EOSs of neutron-star matter, presented
 in Fig.~\ref{fig:eos}.

Stability with
respect to the mass-shedding from the
equator implies that at a given gravitational mass $M$
the circumferential equatorial radius $R_{\rm eq}$
should be smaller than $R_{\rm max}$ which corresponds to the mass shedding
(Keplerian) limit.  The value of $R_{\rm max}$
results from the condition that the frequency of a test
particle at circular equatorial orbit of radius $R_{\rm max}$
just above the equator of the {\it actual rotating star}  is
equal to the rotational frequency of the star.
This condition sets the bound on our rotating configurations from the right
side on $M(R)$ plane (the highest radius and the lowest central density).

The limit for most compact stars (the lowest radius and the highest central density)
is set by the onset of instability with respect to the axisymmetric oscillations
defined by the condition:
 \begin{equation}
 \left({\partial M\over \partial \rho_{\rm c}}
 \right)_J=0~,
 \label{eq:ax-sym.stab.line}
 \end{equation}

For stable configurations we have:
 \begin{equation}
 \left({\partial M\over \partial \rho_{\rm c}}
 \right)_J>0~,
 \label{eq:ax-sym.stab}
 \end{equation}
 %

\section{Neutron stars at 1122 Hz}
\label{sect:NS}

In this section we present the parameters of the stellar configurations
rotating at frequency 1122~Hz. For details and discussion see Bejger et al. (2007)\cite{BHZ2007}
\begin{figure}[h]
\centering
\resizebox{3.5in}{!}{\includegraphics[]{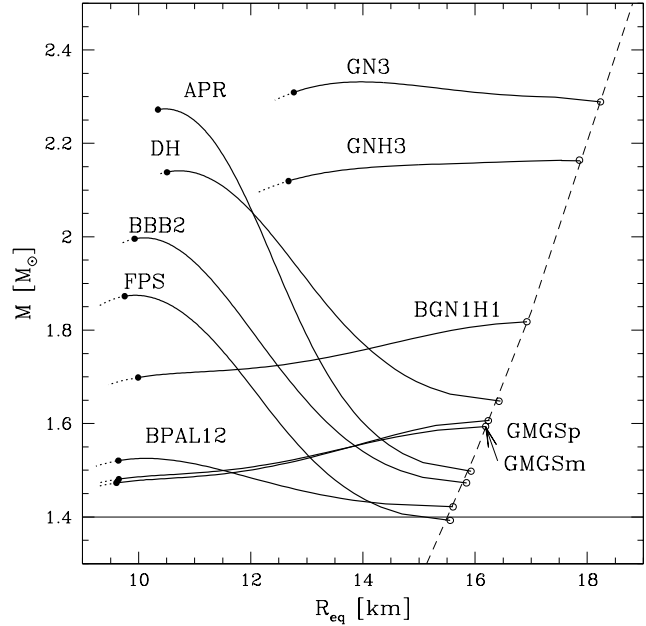}}
\caption{Gravitational mass, $M$, vs. circumferential equatorial radius,
$R_\mathrm{eq}$,    for neutron stars stably rotating at $f=1122$
Hz, for ten EOSs  (Fig.~\ref{fig:eos}).
Small-radius termination by filled circle: setting-in of
instability with respect to the axi-symmetric perturbations.
Dotted segments to the left of the filled circles: configurations unstable with
respect to those perturbations.
 Large-radius termination by an open circle: the mass-shedding
instability. The mass-shedding points are very well fitted by the
dashed curve  $R_{\rm min}=15.52\;(M/1.4 M_\odot)^{1/3}\;$km.
For further explanation see the text.
 }
\label{fig:MR-NS}
\end{figure}

  It is interesting that the relation between the calculated values
  of $M$ and $R_{\rm eq}$ at the "mass shedding point" is extremely
well approximated by the formula for the orbital frequency for
a test particle  orbiting at  $r=R_{\rm eq}$ in the
Schwarzschild space-time of a {\it spherical mass} $M$
(which can be replaced by a point mass $M$ at $r=0$). We denote
the orbital frequency of such a test particle by
$f^{\rm Schw.}_{\rm orb}(M,R_{\rm eq})$.
The formula giving the locus of points satisfying
$f^{\rm Schw.}_{\rm orb}(M,R_{\rm eq})=1122\;$Hz, represented by a dash
line in Fig.\ \ref{fig:MR-NS}, is
\begin{equation}
{1\over 2\pi}\left( {GM\over{ R_{\rm eq}}^3}\right)^{1/2}=1122~{\rm
Hz}~.
\label{eq:f-orb.1122Hz}
\end{equation}
This formula for the Schwarzschild metric coincides with that
obtained in Newtonian gravity for a point mass $M$. It
passes through (or extremely close to)  the open circles
denoting the actual mass shedding (Keplerian) configurations.
This  is  quite remarkable
in view of rapid rotation and strong flattening of neutron star
at the mass-shedding point.
Equation (\ref{eq:f-orb.1122Hz}) implies
\begin{equation}
R_{\rm max}=15.52\;\left({M\over 1.4\;M_\odot}\right)^{1/3}\;{\rm km}~.
\label{eq:Rmax.1122Hz}
\end{equation}

\section{Submillisecond pulsars}

In this section we present results for neutron stars rotating at
submillisecond periods for a broad range of frequencies
(1000 - 1600 Hz)

\begin{figure}
\psfig{file=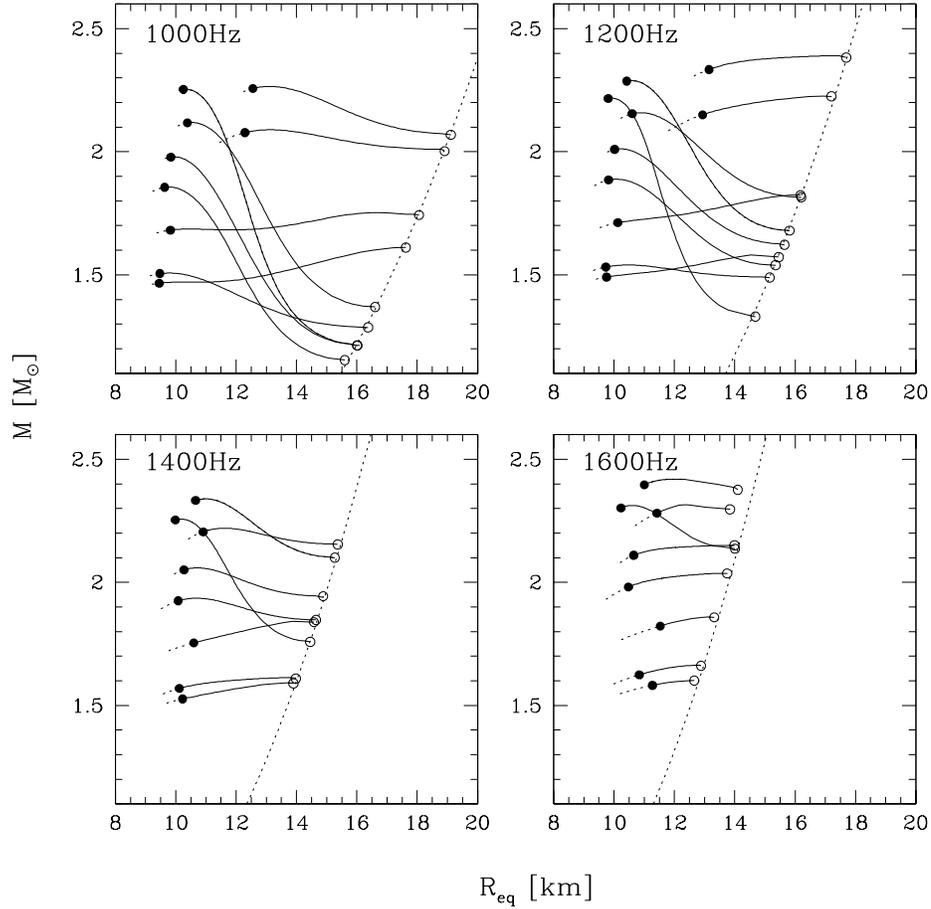,width=5in}
\caption{Mass vs radius relation for submillisecond neutron stars.}
\label{fig:MRsubms}
\end{figure}

In Fig.\ref{fig:MRsubms} we presented mass vs radius relation
for very fast rotating neutron stars. As we see the shape of
the $M(R_{\rm eq})$ curves is more and more flat as rotational
frequency increases. For $f_{\rm rot}=1600~{\rm Hz}$ the
curves $M(R_{\rm eq})$ are almost horizontal and the mass for
each EOS is quite well defined and curves for different models
of dense matter are typically well separated. In principle this
property could be used for selecting  "true EOS" if we detect a
very rapid pulsar and are able to estimate its mass.

In each panel the dotted curve corresponds to the formula (of
which Eq.\;(3) is a special case for $f=1122$ Hz)
\begin{equation}
M={4\pi^2 f^2\over G}R_{\rm eq}^3
\label{eq:M(R)f.ms}
\end{equation}
used for a given frequency (1000~Hz, 1200~Hz,
1400~Hz, 1600~Hz). As we see this formula works perfectly in
very broad range of rotational frequencies (recently this
formula has been tested by Krastev et al.\cite{KLW2007}
for the frequency 716~Hz.)

\section{Accretion}

We studied the mechanism of the spin-up of neutron star due to the accretion from the last
stable orbit (the innermost stable circular orbit - ISCO).
As an example e discuss this subject for DH EOS.

We  calculate the spin-up following the prescription given by
Zdunik et al.\cite{ZHG2002,ZHB2005}.
The value of specific angular momentum per unit baryon mass
of a particle orbiting the neutron star at the ISCO, $l_{\rm IS}$,
is calculated by solving exact equations of the orbital
motion of a particle in the space-time produced by a rotating
neutron star, given in Appendix A of Zdunik et al.\cite{ZHG2002}.

Accretion of an infinitesimal amount of baryon mass ${\rm
d}M_{\rm B}$ onto a rotating neutron star is assumed to lead
to a new quasi stationary rigidly rotating configuration of
mass $M_{\rm B}+{\rm d}M_{\rm B}$ and angular momentum $J+{\rm
d}J$, with
\begin{equation}
{\rm d}J= x_{l}l_{\rm IS} \;{\rm d}M_{\rm B}~,
\label{eq:lms}
\end{equation}
where  $x_l$ denotes the fraction of the angular momentum of
the matter element transferred to the star. The remaining
fraction $1-x_l$ is assumed to be lost via radiation or  other
dissipative  processes.

\begin{figure}
\center{
\psfig{file=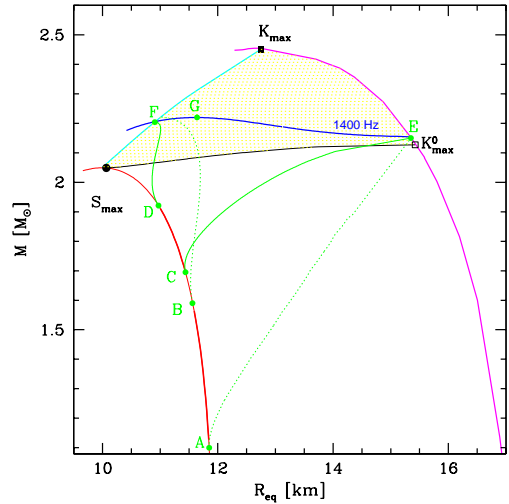, width=2.8in}
}
\caption{Mass vs radius relation for accreting NS with DH EOS. The example for $f_{\rm rot}=1400$~Hz}
\label{fig:MR1400}
\end{figure}

We present results for two choices of $x_l$: $x_l=1$ and
$x_l=0.5$ when all or half of the angular momentum of the
accreting matter is transferred to the star from the ISCO. In
Fig. \ref{fig:MR1400} we plot the curve $M(R)$ corresponding
to the frequency of rotation $f_{\rm rot}=1400$~Hz for DH EOS.
Point F on this curve corresponds to the onset of instability
with respect to axi-symmetric oscillations (condition given by
Eq.\ (\ref{eq:ax-sym.stab.line})). Point E is the Keplerian
configuration at frequency 1400 Hz, and G - corresponds to
maximum mass along the curve with fixed frequency.

The curves starting at points A,B,C and D are the track of the accreting neutron star defined by the
Eq.\ ( \ref{eq:lms})  for $x_l=1$  (solid line) and $x_l=0.5$ (dotted line) for cases C,D and A,B respectively.
To reach the configuration rotating at frequency 1400~Hz we have to start with the nonrotating neutron star
between the points C and D (if $x_l=1$) of A and B  (if $x_l=0.5$)
As we see the actual frequency of rotation sets the limits on the initial mass of nonrotating star,
which can be spun-up to this frequency due to the accretion. For $f_{\rm rot}=1400$~Hz these limits in the
case $x_l=1$ are: $1.7\msol<M_i<1.92\msol$.

\begin{figure}
\center{
\psfig{file=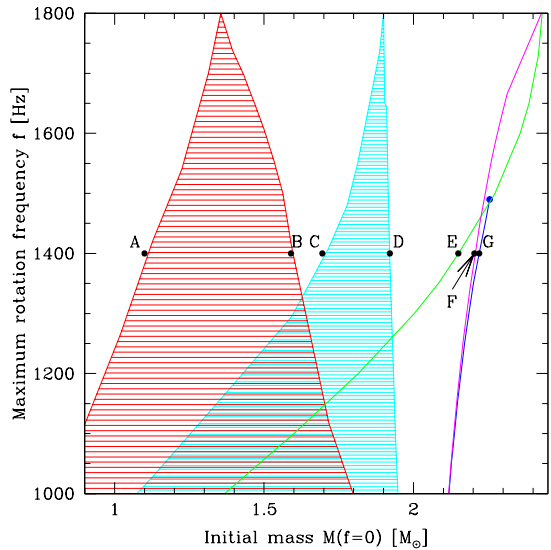, width=3in}}
\caption{(Color online) The
limits for the initial mass of accreting neutron star to be
spun-up up to the given frequency. The labeled points
correspond to the points in Fig. \ref{fig:MRsubms}. The region
to the left (red) corresponds to $x_l=0.5$ and the central
region (cyan) to $x_l=1$. The non-shaded region (right)
correspond to the allowable final masses.} \label{fig:fmacc}
\end{figure}

In Fig  \ref{fig:fmacc} we plotted for DH EOS these limits for the initial mass (non-rotating)
of accreting NS provided that
this star could be spun-up to the given frequency. We presented results for two assumed values of the
parameter $x_l$.
The shaded area shows the allowed initial masses of nonrotating star to reach by the accretion a
given rotational
frequency. Left "triangle" corresponds to $x_l=0.5$, the right shaded triangle (with points C and D) to $x_l=1$

The curves on the right represent  the
limits on the actual mass of rotating NS. The three curves correspond to the
location of the three points E,F, G in Fig \ref{fig:MRsubms}. The curve with the point E (green) is defined
by the Keplerian frequency of rotating star, the point F (and magenta line) correspond to the
boundary resulting from the instability with respect to axi-symmetric perturbations. The point G is the mass at the maximum point of the curve
with fixed frequency (1400~Hz). As the frequency increases the mass at Keplerian point increases more rapidly than that
defined by the onset of instability at the maximum mass (points F and G respectively). For high frequencies the maximum mass
of the stars rotating at fixed frequency is given by the value for Keplerian configuration.
For DH EOS at frequency $\simeq 1500$~Hz the point G disappears and for faster rotation the curve $M(R_{\rm eq})$
 monotonically
increases.
The region of the masses of the stars rotating at high frequency is very narrow.
For $f_{\rm rot} > 1400$~Hz it is smaller than
$0.1\msol$ (see also discussion of Fig.\ref{fig:MRsubms})

\section{Discussion and conclusions}

The $M(R_{\rm eq})$ curve for $f\gtrsim 1400$ Hz is flat. Therefore,
for given EOS the mass of NS is quite well defined.
Conversely, measured mass of a NS rotating at $f\gtrsim 1400$ Hz will
allow us to unveil the actual EOS. The "Newtonian" formula for
the Keplerian frequency works surprisingly well for precise
2-D simulations and sets a firm upper limit on $R_{\rm eq}$
for a given $f$. Finally, observation of $f\gtrsim 1200$ Hz  sets
stringent limits on the initial mass of the nonrotating star
which was spun up to this frequency by accretion.
\label{sect:discuss}
%

%
\section*{Acknowledgments}
 This work was partially
supported by the Polish MNiSW grant no. N203.006.32/0450 
and by the LEA
Astrophysics Poland-France
(Astro-PF) program. MB was
also partially supported by the Marie Curie Intra-european
Fellowship MEIF-CT-2005-023644.

\end{document}